\documentclass[9pt,twocolumn,twoside]{pnas-new}
\usepackage{natbib}
\usepackage{comment}

\templatetype{pnasresearcharticle} 

\setboolean{displaywatermark}{false}

\title{\texttt{APACE}: \texttt{AlphaFold2} and advanced computing as a service for accelerated discovery in biophysics}

\author[a,b,c]{Hyun Park}
\author[a,d,e]{Parth Patel}
\author[e]{Roland Haas}
\author[a,f,g]{E.~A. Huerta}

\affil[a]{Data Science and Learning Division, Argonne National Laboratory, Lemont, Illinois 60439, USA}
\affil[b]{Theoretical and Computational Biophysics Group, Beckman Institute for Advanced Science and Technology, University of Illinois at Urbana-Champaign, Urbana, Illinois 61801, USA}
\affil[c]{Biophysics and Quantitative Biology, University of Illinois at Urbana-Champaign, Urbana, Illinois 61801, USA}
\affil[d]{Department of Computer Science, University of Illinois at Urbana-Champaign, Urbana, Illinois 61801, US}
\affil[e]{National Center for Supercomputing Applications, University of Illinois at Urbana-Champaign, Urbana, Illinois 61801, USA}
\affil[f]{Department of Computer Science, The University of Chicago, Chicago, Illinois 60637, USA}
\affil[g]{Department of Physics, University of Illinois at Urbana-Champaign, Urbana, Illinois 61801, USA}

\leadauthor{Huerta}

\significancestatement{We introduce \texttt{APACE}, 
\underline{A}l\underline{p}haFold2 and 
\underline{a}dvanced \underline{c}omputing as a 
s\underline{e}rvice, a novel computational framework 
that optimizes \texttt{AlphaFold2} to run at scale 
in high performance computing platforms, and 
which effectively handles this TB-size AI model 
and database. We showcase the use of \texttt{APACE} 
in the Delta and Polaris supercomputers 
to accelerate protein 
structure prediction for a variety of 
proteins, and demonstrate that using 
200 ensembles distributed over 300 NVIDIA A100 GPUs, 
\texttt{APACE} reduces time-to-insight from days to minutes. This framework may be readily linked with 
self-driving laboratories to enable 
automated discovery at scale.}

\authorcontributions{E.A.H. designed research; H.P., P.P., and R.H. performed research; H.P., P.P., and 
E.A.H. contributed new reagents/analytic tools; H.P., P.P., and E.A.H. analyzed data; and H.P., P.P., 
and E.A.H. wrote the paper.}

\authordeclaration{The authors declare no competing interests.}

\correspondingauthor{\textsuperscript{1}To whom correspondence should be addressed. Eliu Huerta, e-mail: elihu@anl.gov}


\begin{abstract}
The prediction of protein 3D 
structure from amino acid sequence is a computational 
grand challenge in biophysics, and plays a key role in 
robust protein structure prediction algorithms, 
from drug discovery to genome interpretation. 
The advent of AI models, such as 
\texttt{AlphaFold}, is revolutionizing 
applications that depend on robust protein structure prediction algorithms. To maximize the impact, 
and ease the 
usability, of these novel AI tools 
we introduce \texttt{APACE}, 
\underline{A}l\underline{p}haFold2 and 
\underline{a}dvanced \underline{c}omputing as a 
s\underline{e}rvice, a novel computational framework  
that effectively handles this AI model 
and its TB-size database to conduct accelerated protein 
structure prediction analyses in modern supercomputing 
environments. We deployed \texttt{APACE} 
in the Delta and Polaris supercomputers, and quantified 
its performance for accurate protein structure predictions using four exemplar proteins: 6AWO, 
6OAN, 7MEZ, and 6D6U. Using up to 300 ensembles, 
distributed across 200 NVIDIA A100 GPUs, we found 
that \texttt{APACE} 
is up to two orders of magnitude faster than 
off-the-self \texttt{AlphaFold2} implementations, 
reducing time-to-solution from weeks to minutes. This 
computational approach may be readily linked with 
robotics laboratories to automate and accelerate 
scientific discovery. 
\end{abstract}

\dates{This manuscript was compiled on \today}
\doi{\url{www.pnas.org/cgi/doi/10.1073/pnas.XXXXXXXXXX}}

\begin{document}

\maketitle
\thispagestyle{firststyle}
\ifthenelse{\boolean{shortarticle}}{\ifthenelse{\boolean{singlecolumn}}{\abscontentformatted}{\abscontent}}{}

\dropcap{I}nnovation at the interface of artificial intelligence (AI) and advanced computing is enabling breakthroughs in science and engineering~\cite{DeepLearning,Nat_Rev_2019_Huerta,2022NatRP...4..761K,sci_miw,BIANCHINI2022104604,Crawford+2021,DeepLearning,dl2010review}. The rise of AI models such as 
GPT-4~\cite{openai2023gpt4}, 
\texttt{AlphaFold}~\cite{jumper2021highly}, among others, provides new capabilities to accelerate and automate scientific discovery. However, some of these models have not been 
released to the public, breaking a strong tradition 
in the AI community. It's been argued that the 
sheer size of these AI models prohibits their 
use by a large cross section of potential users. 

To address this shortcoming,  
we demonstrate how to combine large AI models 
with high performance computing platforms to 
empower a broad cross section of users to 
fully exploit the capabilities of AI for scientific 
discovery. We have selected 
\texttt{AlphaFold2}~\cite{bryant2022improved} as the science driver for 
this study, since this AI model is revolutionizing 
discovery in biophysics, and its use 
for accurate and rapid protein structure prediction (PSP) 
demands an optimal use of modern supercomputing 
environments. Here, we demonstrate how to optimize 
\texttt{AlphaFold2} and its database, which exceed 
2.6TB in data storage, to reduce the time needed for 
accurate PSPs from weeks to minutes.

\paragraph{\texttt{AlphaFold2}'s features.}  
With a deep learning technique based structure prediction, \texttt{AlphaFold2}~\cite{jumper2021highly} showed 
unprecedented performance at the 14th Community Wide Experiment on the Critical Assessment of Techniques for Protein Structure Prediction (CASP14)~\cite{moult2020critical}, later improved further with multimer prediction \cite{evans2021protein}. Ever since, efforts to make \texttt{AlphaFold2} faster~\cite{zhong2022parafold,mirdita2022colabfold}, to predict protein complex, e.g., antibody~\cite{yin2022benchmarking, bryant2022improved}, and to sample diverse protein conformations~\cite{
 meller2023accelerating, faezov2023alphafold2, sala2023modeling} have come to fruition. In this study we use \texttt{AlphaFold2} version 2.3.0., 
as of August, 2023. The pre-trained neural network 
parameters used include both monomer and multimer v3.

\texttt{AlphaFold2} utilizes Central Processing Units 
(CPUs) to compute key input features: multiple sequence alignment (MSA), and structural templates. 
MSA represents a collection of protein sequence homologues related to the query protein. MSA captures evolutionary relationships between various proteins such as conserved and variation amino acid residues. MSA is computed using CPU-based sequence alignment algorithms, such as \texttt{Jackhmmer}, which align a query protein sequence with known sequence homologues obtained from databases such as Uniclust. \texttt{AlphaFold2} can glean key residue interactions from MSA~\cite{jumper2021highly}.
Furthermore, structural templates refer to experimentally known protein homologue structures that share significant sequence similarity with the query protein. These templates structures are used to improve the accuracy of \texttt{AlphaFold2}'s predictions. CPU-based algorithms, such as \texttt{HHsearch} (for monomer) and \texttt{Hmmsearch} (for multimer), search public protein structure databases such as the Protein Data Bank (PDB). Then, \texttt{AlphaFold2} extracts spatial information from the template relevant to the query protein~\cite{jumper2021highly}.

In the Graphics Processing Unit (GPU) phase, \texttt{AlphaFold2} utilizes the features generated from MSA and templates, passing them through the evoformer network. Evoformer refines representations for both the MSA and pair interactions while iteratively exchanging information between them in a criss-cross fashion to extract amino acid residue relationships. The updated representations then enter a structure module, where predictions for rotations and translations are made to position each residue~\cite{jumper2021highly}.

The resulting predicted 3D structure undergoes a relaxation process via minimization by MD engine to enhance accuracy. Upon generating the final structure, the information cycles back to the beginning of the evoformer blocks in a \textit{recycle} procedure, further refining the structure predictions. Overall, \texttt{AlphaFold2} is trained end-to-end, leading to remarkable accuracy and reliability in predicting protein 3D structures~\cite{jumper2021highly}.

\paragraph{\texttt{APACE}'s improvement over \texttt{AlphaFold2}.} We introduce \texttt{APACE}, 
\underline{A}l\underline{p}haFold2 and 
\underline{a}dvanced \underline{c}omputing as a 
s\underline{e}rvice, 
a computational framework to accelerate 
\texttt{AlphaFold2} through CPU \& GPU optimizations, 
and distributed computing in supercomputing 
environments. Key features of this approach encompass:

\noindent \textbf{Data management.} First, \texttt{APACE} facilitates the usage 
of \texttt{AlphaFold2}'s 2.6 TB AI model and 
database~\cite{jumper2021highly} by hosting it 
at the Delta and Polaris supercomputers~\cite{towns2014xsede}. 
\texttt{AlphaFold2}'s neural networks can readily 
access data by leveraging solid state drive (SSD) data storage, and Infinite Memory Engine (IME) data staging.

\noindent \textbf{CPU optimization.} Second, inspired 
by~\cite{zhong2022parafold}, \texttt{APACE} uses Ray library's~\cite{moritz2018ray} CPU optimization to parallelize CPU intensive MSA and template computation calculations. As part of CPU optimization, \texttt{APACE} allocates higher CPU cores to MSA/template search tools, rather than default numbers (4 or 8), which showed heuristic speed improvement in our experiments. In addition, in a similar manner as~\cite{zhong2022parafold}, we also implemented a checkpoint to circumvent redundant MSA/template steps 
if \textit{features.pkl} file exists, i.e., an intermediate file storing MSA/template search result.

\noindent \textbf{GPU optimization.} Third, \texttt{APACE} uses Ray library's~\cite{moritz2018ray} GPU optimization to parallelize GPU intensive neural network protein structure prediction steps. An important key difference from ParaFold~\cite{zhong2022parafold} in terms of GPU speedup is that, ParaFold predicted one conformation with only \textit{model\_1} (a template based pretrained model) mostly on peptide sequences (e.g., with an average size less then 100 amino acid residues), rather than protein sequences (i.e., $\sim 400$ amino acid residues and more). In stark 
contrast, \texttt{APACE} predicts multiple protein conformations for each protein sequence, and peptide sequence if necessary, with all five pretrained models 
in parallel, which is computationally demanding.

\noindent \textbf{New functionalities.} Fourth, \texttt{APACE} can predict multiple monomer conformations per pretrained neural network model (out of five models), a simple functionality existing only in multimer prediction in the original \texttt{AlphaFold2} model. \texttt{APACE} includes functionalities such as enabling dropout during structure prediction, changing number of Evoformer~\cite{jumper2021highly} recycles, or subsampling MSA options, as provided in~\cite{mirdita2022colabfold}.

\section*{Results and Discussions}
\label{rnd}

We completed three computational experiments to carry out a detailed comparison 
between \texttt{APACE} and the original \texttt{AlphaFold2} 
model. We describe each experiment at a time, and then provide the corresponding results. 

These results were obtained using the Delta and Polaris supercomputers, housed at the National Center for 
Supercomputing Applications, and at 
the Argonne Leadership Computing Facility (ALCF), respectively. 
Both machines provide highly capable GPU-focused compute environment for GPU and CPU workloads. 

Delta offers a mix of standard and reduced precision 
GPU resources, as well as GPU-dense nodes with both NVIDIA and AMD GPUs. It also provides high performance node-local 
SSD scratch file systems, as well as both standard Lustre and relaxed-POSIX parallel file systems 
spanning the entire resource. On the other hand, 
the Polaris supercomputer has 560 nodes. Each compute 
node consists of 1 AMD EPYC Milan processor,
4 NVIDIA A100 GPUs, unified memory architecture, 2 fabric endpoints, and 2 NVMe SSDs. The system interconnect 
is HPE Slingshot 11, and uses a Dragonfly topology 
with adaptive routing.
 
We compared \texttt{APACE} and \texttt{AlphaFold2} performance 
using both NVIDIA A100 and A40 GPUs in Delta, and NVIDIA A100 GPUs 
in Polaris. The computational benchmarks we report below in 
terms of 
CPU and GPU runtimes were 
extracted from the generated timings.json file 
of both \texttt{APACE} and \texttt{AlphaFold2}.

\paragraph{Experiment 1: Predicting structures for 4 benchmark proteins.} 
Four proteins were selected as benchmarks to assess the effectiveness and operational proficiency of \texttt{APACE}. To predict protein structures 
with \texttt{APACE}, we developed scientific software that enables 
users to provide suitable headers in \textit{sbatch} scripts, and to 
load the appropriate environment and module, that are 
used to successfully submit and complete simulations in the Delta 
and Polaris supercomputers. These are the SLURM parameters we used: \textit{-{}-mem=240g, -{}-nodes=10,
-{}-exclusive, -{}-ntasks-per-node=1, -{}-cpus-per-task=64, -{}-gpus-per-task=4, -{}-gpus-per-node=4}. The neural network and MSA/template related parameters were the same as \texttt{AlphaFold2}.

\noindent \textbf{Monomers.} We used the monomer protein 6AWO (serotonin transporter) as a basic structure to test baseline prediction 
accuracy and conformational diversity using a total of five models. 
Thus, we created a Ray cluster consisting of 8 NVIDIA A100/A40 GPUs (equivalent to 2 A100/A40 GPU 
nodes in Delta and Polaris) to 
facilitate both CPU and GPU parallel execution and relaxation for 
all 5 models, i.e., 1 structure per model, as in the case of \texttt{AlphaFold2}.

\begin{table*}[!htbp]
\centering
\caption{Performance benchmarks between off-the-shelf \texttt{AlphaFold2} and our {\cal{\texttt{APACE}}} CPU \& GPU optimized framework for four exemplar proteins. We 
present results for two types of GPUs available in the 
Delta supercomputer, NVIDIA A40 and A100 GPUs. We also present results using the Polaris supercomputer housed at the Argonne Leadership Computing Facility. }
\begin{tabular}{lrrrrrrr}
Protein & \# of  & \texttt{AlphaFold2}-A40 & \texttt{AlphaFold2}-A100 & {\cal{\texttt{APACE}}}-A40 & {\cal{\texttt{APACE}}}-A100 & \texttt{AlphaFold2}-Polaris & {\cal{\texttt{APACE}}}-A100 Polaris \\
& ensembles & CPU/GPU [min] & CPU/GPU [min] & CPU/GPU [min] & CPU/GPU [min] & CPU/GPU [min] & CPU/GPU [min]\\
\midrule
1. 6AWO & 5 & 33.0 / 17.7 & 33.0 / 12.8 & 16.1 / 4.0 & 16.1 / 2.9 & 189.0 / 46.8 & 92.2 / 9.4\\
2. 6OAN & 40 & 99.1 / 268.1 & 99.4 / 181.4 & 56.5 / 7.9 & 57.1 / 5.6 & 306.2 / 593.5 & 175.9 / 14.8\\
3. 7MEZ & 40 & 100.7 / 3756.3 & 100.7 / 2339.4 & 58.0 / 100.1 & 58.8 / 63.0 & 556.2 / 3640.9 & 324.8 / 91.0\\
4. 6D6U & 40 & 143.6 / 1528.7 & 143.8 / 786.9 & 89.3 / 72.2 & 89.4 / 35.1 & 485.9 / 1279.1 & 302.1 / 32.0\\
\bottomrule
\label{tb1}
\end{tabular}
\end{table*}

\noindent \textbf{Multimers.} For multimer proteins, we tested 
6OAN, Duffy-binding protein bound with single-chain variable fragment antibody~\cite{yin2022benchmarking}; 7MEZ, phosphoinositide 3-kinase~\cite{bryant2022improved}; and 6D6U, a 3 distinct chain heteropentamer GABA transporter, which represents a more challenging case for multimer prediction. For each of these proteins, we had 8 structure predictions per model, yielding a total of 40 predictions (5 ensemble modes x 8 predictions per model). To facilitate concurrent execution and relaxation for the entire array of 40 models, 40 NVIDIA A100 and also 40 A40 GPUs (10 A100/A40 GPU nodes) were harnessed using a Ray cluster.

To initiate a Ray cluster utilizing compute nodes 
(as described in Methods), we first fetched a list of available compute nodes and their IP addresses. 
We then launched a head Ray process using one 
of these nodes, referred to as the ``head node''. Subsequently, we started Ray worker processes for the remaining compute nodes. Each worker is equipped with all 4 GPUs, and Ray automatically determines the utilization of available GPUs for running and relaxation of models. The workers are then linked to the head node by providing the head node's address.

We utilize \textit{srun} via message passing interface (MPI) to start the workers on the compute nodes. This is necessary because the \textit{sbatch} script executes solely on the first compute node. 
Given the simultaneous launch of all Ray processes using MPI, we incorporate safeguards to prevent race conditions. The race condition safeguards ensure that the head node is started before the worker nodes and the beginning of predictions.

After the underlying Ray cluster was ready, we established a connection to it using \textit{ray.init} within the \textit{run\_alphafold.py} code and initiated the prediction of the protein structure using \texttt{APACE}. Ray automatically allocates resources and concurrently executes MSA tools on CPUs, model runs, and model relaxation on a distinct GPU. This operation efficiently harnesses the full computational potential of both CPU cores and GPUs available on the compute node.

\paragraph{Experiment 1: Results and Discussion.}

\noindent \paragraph{CPU acceleration.} Through the 
implementation of parallel optimization techniques, 
\texttt{APACE} achieved an 1.8X average CPU speedup in 
Delta, and 1.78X average CPU speedup 
in Polaris. These results 
are independent of the number of compute nodes.

\noindent \paragraph{GPU acceleration.} \texttt{APACE} achieves significant GPU speedups. The following 
results were obtained using 8 GPUs for 6AWO, and 
40 GPUs for 6OAN, 7MEZ, and 6D6U:
\begin{enumerate}[nosep]
    \item 6AWO. 4.4X speedup on both A40 and A100 GPUs for Delta; and 4.98X speedup on Polaris.
    \item 6OAN. 34X and 32.4X speedup on A40 and A100 GPUs, respectively, for Delta; and 40.1X speedup for Polaris.
    \item 7MEZ. 37.5X and 37.1X speedup on A40 and A100 GPUs, respectively, for Delta; and 40X speedup for Polaris.
    \item 6D6U. 21.2X and 22X speedup on A40 and A100 
    GPUs, respectively, for Delta; and 40X speedup 
    for Polaris.
\end{enumerate}

\noindent We summarize these results in \autoref{tb1}. We also note that prediction times are consistently shorter when using NVIDIA A100 GPUs. In brief, \texttt{APACE} provides remarkable speedups for 
basic and complex structures, retaining the accuracy and 
robustness of the original \texttt{AlphaFold2} model.
Furthermore, \texttt{APACE} can readily be used for analyses at 
scale using hundreds of GPUs, as shown below.

\paragraph{Experiment 2: Predicting Protein 7MEZ using 100 and 200 NVIDIA A100 GPUs.} To quantify 
the performance and scalability of \texttt{APACE} in the Delta 
and Polaris supercomputers, we conducted protein 7MEZ predictions utilizing a significant number of compute nodes. Specifically, we utilized 100 NVIDIA A100 GPUs, which correspond to 25 A100 GPU compute nodes to generate predictions (20 predictions per model). Likewise, we leveraged the computational power of 200 NVIDIA A100 GPUs, equivalent to 50 A100 compute nodes to generate a total of 200 predictions (40 predictions per model). To predict the structures, the \textit{sbatch} script was modified 
to allocate the correct number of compute nodes. We also modified 
the srun and singularity run parameters to successfully complete 
these calculations.

\paragraph{Experiment 2: Results and Discussion.}
\texttt{APACE} delivered remarkable 
speedups. If we compute 100 ensembles 
(distributed over 100 GPUs) for protein 7MEZ, \texttt{APACE} 
completed the required calculations within 67.8 minutes, as 
opposed to \texttt{AlphaFold2}'s 6068.8 minutes 
(101.1 hours/4.2 days) in Delta. In 
Polaris we observe that \texttt{APACE}  reduced 
time-to-solution from 8793.3 minutes (146.5 hours/6.1 days) 
to 87.9 minutes.

Similarly, if we now require 
200 ensembles for the same protein, \texttt{APACE} 
in Delta completed all predictions within 64 minutes, as 
opposed to the 12023.3 minutes (200.4 hours/8.3 days) that 
would be needed using the original \texttt{AlphaFold2} method. 
In Polaris, \texttt{APACE} 
reduced time-so-solution from 12741.2 minutes 
(212.4 hours/8.8 days) to only 84.9 minutes.

Finally, using 300 ensembles for protein 7MEZ, 
\texttt{APACE} in Delta completed all predictions within 68.2 minutes, as 
opposed to the 18064.3 minutes (301.1 hours/12.5 days) that 
would be needed using the original \texttt{AlphaFold2} method. 
In Polaris, \texttt{APACE} 
reduced time-so-solution from 15295.6 minutes 
(254.9 hours/10.6 days) to only 76.9 minutes. 
These results are summarized in \autoref{tb2}.

\begin{table*}[!htbp]
\centering
\caption{Performance benchmarks between off-the-shelf \texttt{AlphaFold2} and {\cal{\texttt{APACE}}} for protein 7MEZ. We present results for 25, 50, and 75 nodes in Delta and the Polaris supercomputers. Each node has 4 NVIDIA A100 GPU.}
\begin{tabular}{lrrrr}
Nodes / Ensembles & \texttt{AlphaFold2}-Delta  & {\cal{\texttt{APACE}}}-Delta  & \texttt{AlphaFold2}-Polaris & {\cal{\texttt{APACE}}}-Polaris \\
& CPU/GPU [min] & CPU/GPU [min] & CPU/GPU [min] & CPU/GPU [min] \\
\midrule
25 / 100  & 100.7 / 6068.8 & 58.8 / 67.8 & 556.2 / 8793.3 & 324.8 / 87.9 \\
50 / 200 & 100.7 / 12023.3 & 58.8 / 64.0 & 556.2 / 12741.2 & 324.8 / 84.9 \\
75 / 300 & 100.7 / 18064.3 & 58.8 / 68.2 & 556.2 / 15295.6 & 324.8 / 76.9 \\
\bottomrule
\label{tb2}
\end{tabular}
\end{table*}

\paragraph{Experiment 3: Ensemble diversity of \texttt{APACE}.} \texttt{AlphaFold2}'s inherent limitations restrict us to generating merely 5 predictions per monomer, e.g., 1 prediction per model, thereby confining the diversity of protein conformation. Moreover, fine-tuning parameters such as dropout remains inaccessible. However, we successfully addressed this constraint by adapting the ColabFold~\cite{mirdita2022colabfold} code. For experimental purposes, we generated 100 structures for protein 6AWO using \textit{-{}-num\_multimer\_predictions\_per\_model=20} while employing the parameter \textit{-{}-use\_dropout=True}. This was accomplished by configuring the \textit{sbatch} script with the appropriate parameters.

\texttt{APACE} enables users to select the 
following options~\cite{mirdita2022colabfold}:

\begin{enumerate}[nosep]
    \item Ensemble of structure module with \textit{-num\_ensemble},
    \item Control for recycles with \textit{-{}-num\_recycles},
    \item Subsampling of MSA with \textit{-{}-max\_seq, -{}-max\_extra\_seq},
    \item Evoformer fusion with \textit{-{}-use\_fuse},
    \item Bfloat16 mixed precision with \textit{-{}-use\_bfloat16},
    \item Bernoulli-masking based diverse conformational sampling with \textit{-{}-use\_dropout}.
\end{enumerate}

\paragraph{Experiment 3: Results and Discussion.}

\begin{figure}[!htbp]
\centering
\centerline{
\includegraphics[width=.4\textwidth]{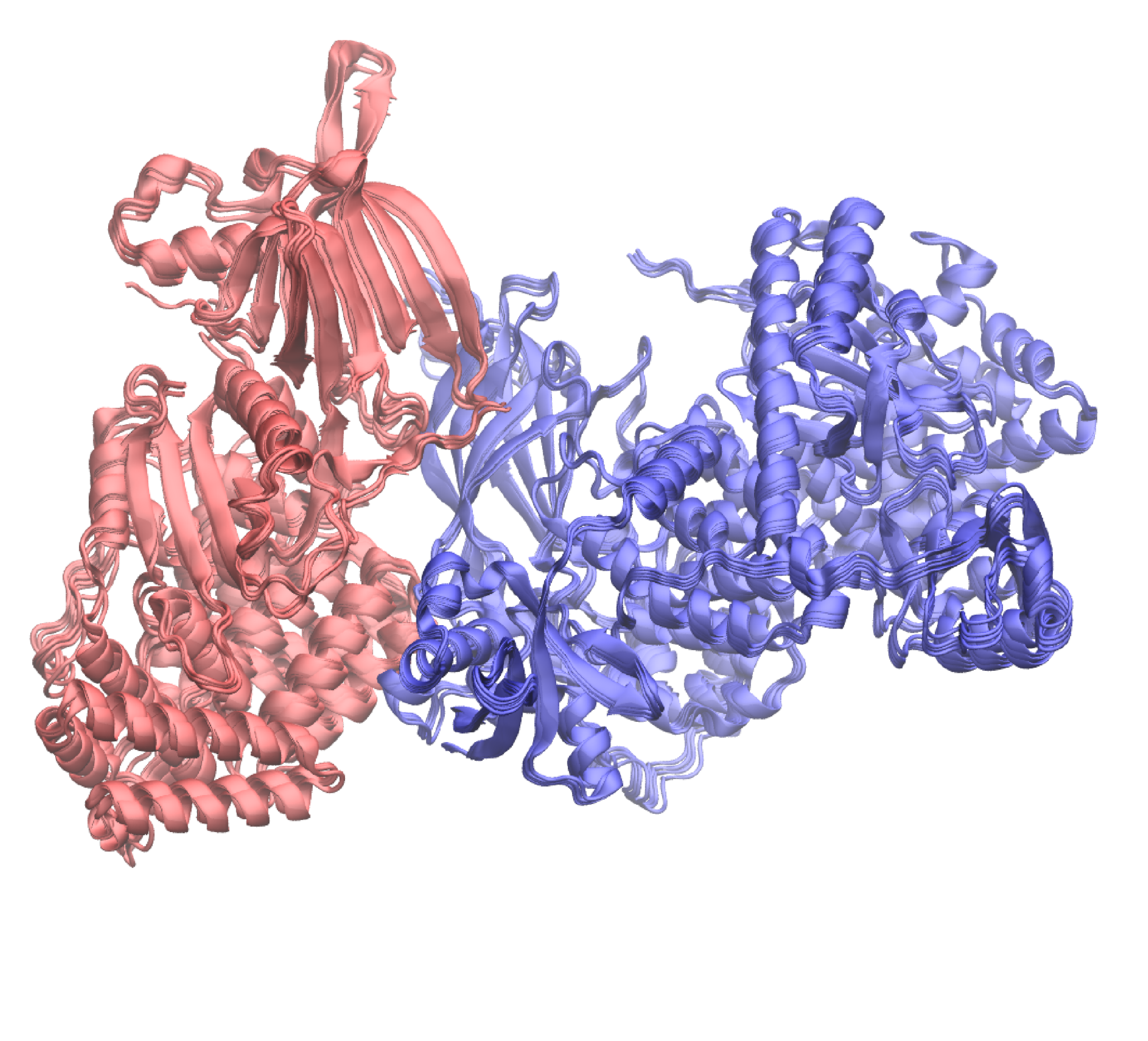}
}
\vspace{-12mm}
\caption{Protein structure used to test 
{\cal{\texttt{APACE}}}: a 
phosphoinositide 3-kinase (PI3K) consisting of 
p110$\gamma$ and p101 subunits (PDB accession: 
7MEZ). The structure has good agreement with ground truth bound structure conformation. Although there are mispredictions of loop secondary structures in p101 (red; top left helical loop; mispredicted as alpha helix rather than loop) subunit, the interface binding pose between p101 and p110$\gamma$ (blue) is well predicted, implying conserved binding interface in evolution. Also, rest of the secondary structures and overall heterodimer structure of the predicted conformational ensemble are comparable with ground truth structure.}
\label{fig:7mez}
\end{figure}

\paragraph{Protein structure prediction and conformational diversity by \texttt{APACE}.}
We have modified \texttt{AlphaFold2} code to mirror ColabFold~\cite{mirdita2022colabfold}'s versatile protein structure prediction pipeline parameter customization. 
With these improvements, we have successfully expanded the spectrum of predictions, thereby enhancing the overall reliability of the predicted structures. Although protein structure prediction is of great significance, we would like to expand \texttt{APACE} to predict conformational diversity since proteins are not static but malleable and flexible structures. Sampling a wide range of conformational ensemble is important for drug discovery~\cite{meller2023accelerating, faezov2023alphafold2, sala2023modeling}.
\begin{figure}[!htbp]
\centering
\centerline{
\includegraphics[width=.5\linewidth]{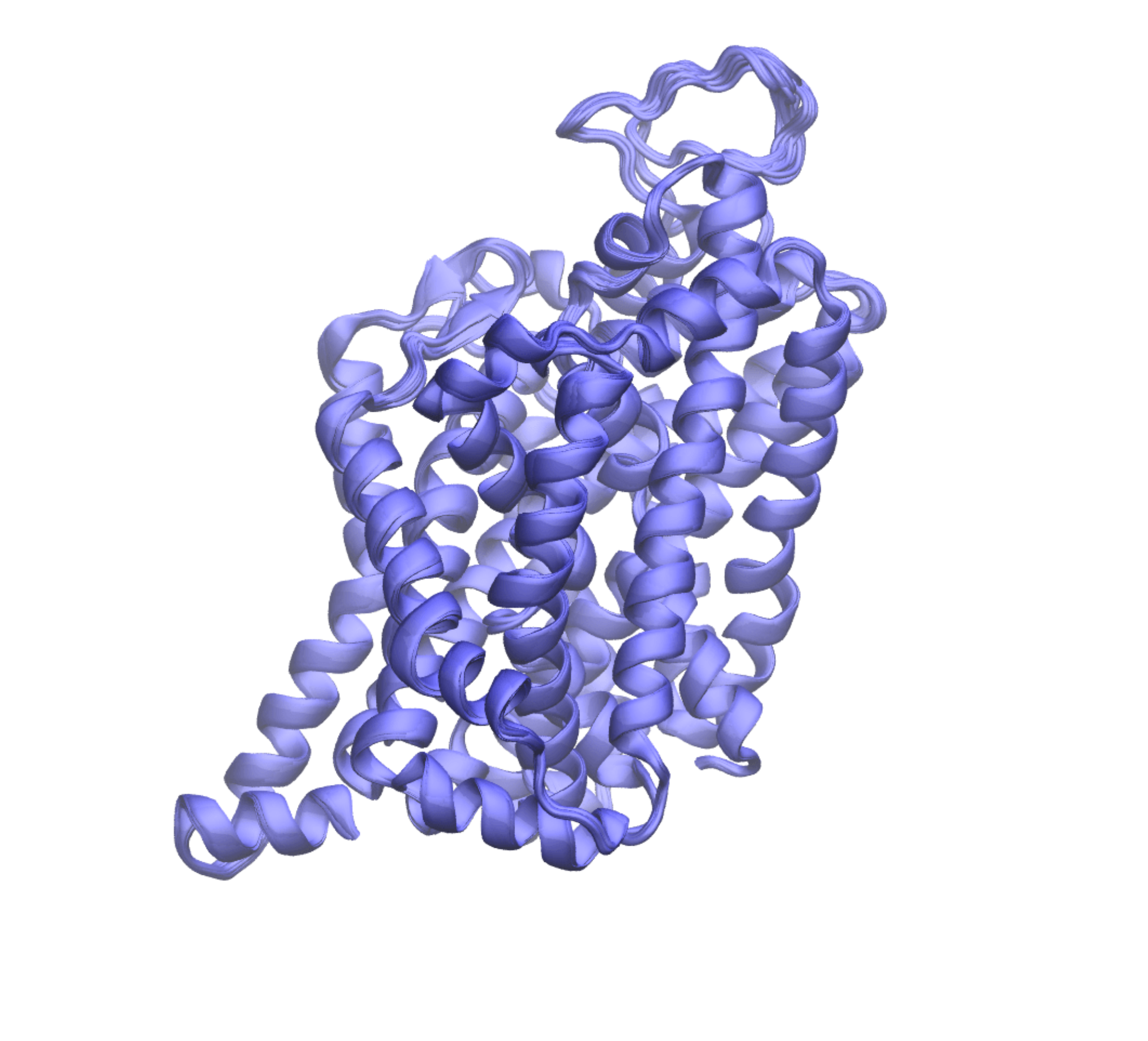}
\includegraphics[width=.5\linewidth]{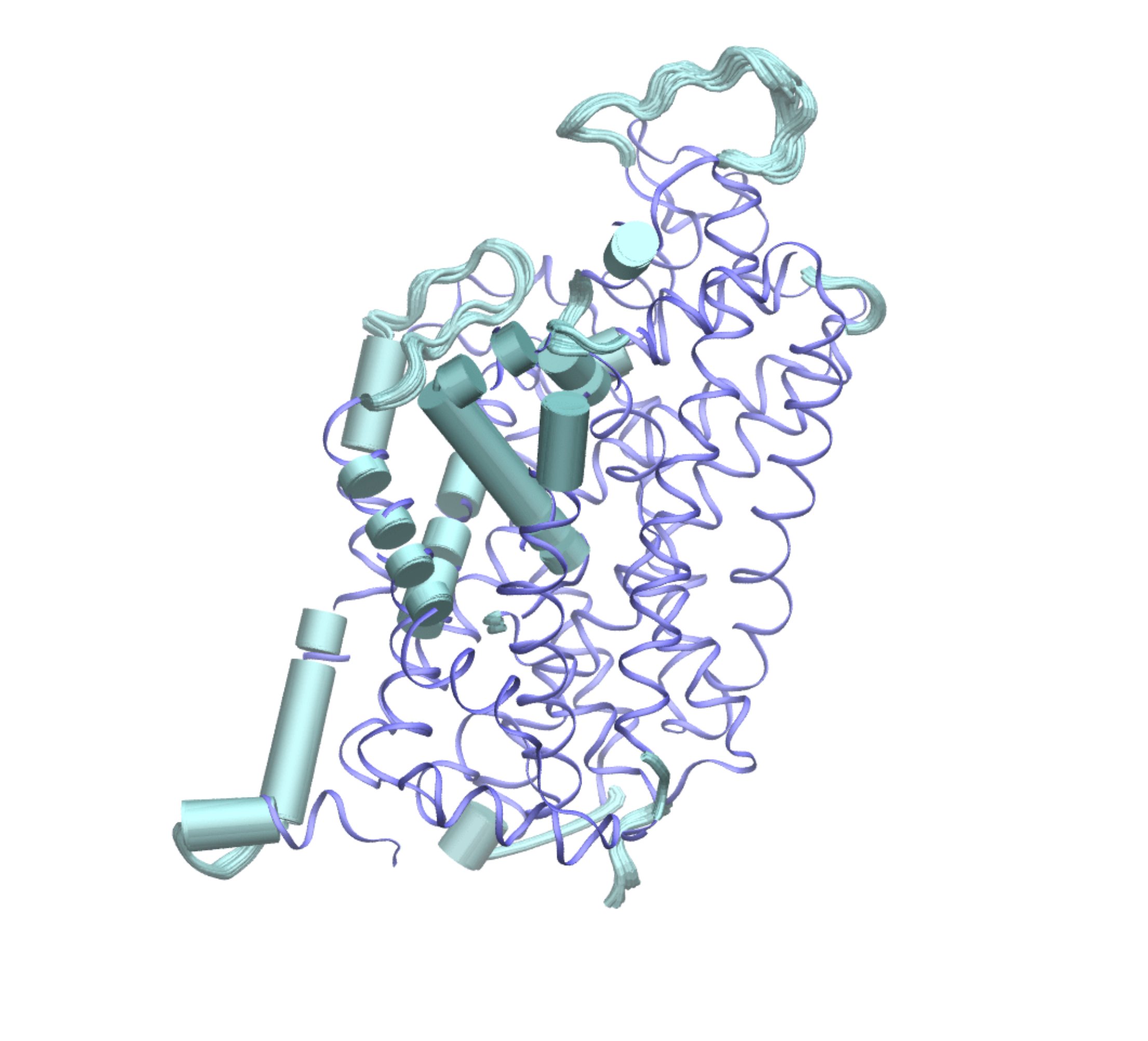}
}
\vspace{-8mm}
\caption{Protein structure used to test 
{\cal{\texttt{APACE}}}: serotonin transporter (PDB 
accession: 6AWO; \textit{shorthand} SERT). Left panel is 100 SERT predicted conformational ensemble overlayed, which has good agreement with ground truth SERT. Right panel is high variant transmembrane domains (TMs), shown in cyan, and computed with root mean square fluctuations (RMSFs) overlayed. Figures are generated with Visual 
Molecular Dynamics (VMD)~\cite{humphrey1996vmd}.}
\label{fig:6awo}
\end{figure}
In the case of 6AWO ($\sim 500$ amino acid residues), \autoref{fig:6awo}, we used our parameter customization 
enhancements (with the option \textit{-{}-use\_dropout=True}) and predicted 100 structures of serotonin transporter (SERT). We have found that the structure predicted by \texttt{APACE} is comparable to the ground truth structure. When we visualize most variant transmembrane domain alpha helices (cyan in right panel), we observe that TM2, TM6, TM10 and TM12 are highlighted. Among these, TM6, TM10 and TM12 are responsible for conformational change or ligand binding from outward-facing to 
inward-facing structures~\cite{coleman2019serotonin,chan2022structural,chan2022substrate}. This implies that \texttt{APACE} learned patterns to predict a wide range of conformational landscape of SERT. Also, given that SERT is a membrane protein, and \texttt{APACE} predicts the conformations without membrane with good accuracy makes the application of \texttt{APACE} promising for drug discovery.   
\begin{figure}[!htbp]
\centering
\centerline{
\includegraphics[width=.8\linewidth]{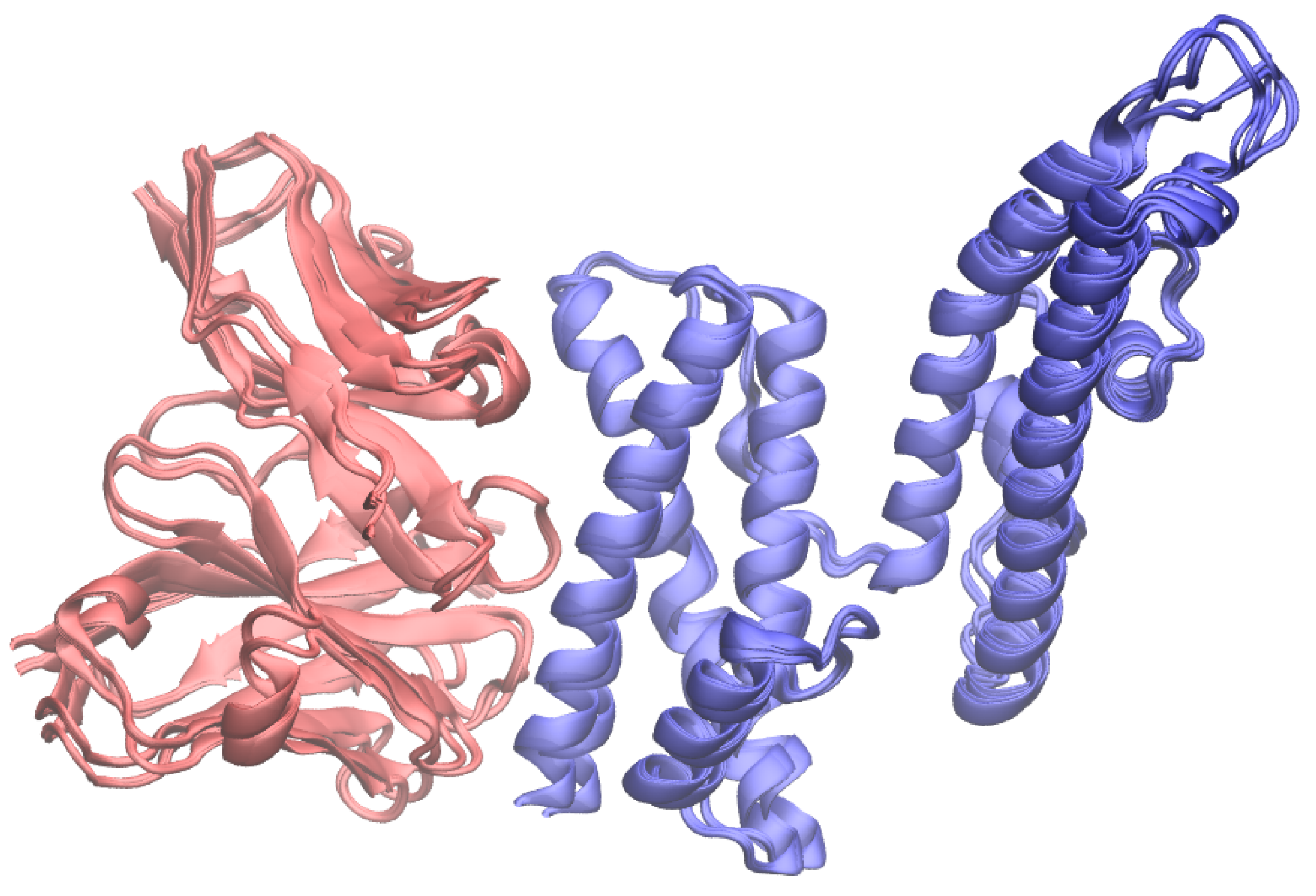}
}
\vspace{-4mm}
\caption{Protein structure used to test 
{\cal{\texttt{APACE}}}: the antibody-antigen 
complex Plasmodium vivax Duffy-binding protein 
(PDB accession: 6OAN). The structure has good agreement with ground truth bound structure conformation. The predicted conformational ensemble of complementary determining region (CDR; loops) of the antibody (red) binding against helical secondary structure epitopes of antibody (blue) are predicted well when compared to ground truth.}
\label{fig:6oan}
\end{figure}
In the case of 6OAN, 7MEZ and 6D6U ($\sim 600, 2000, 1800$ amino acid residues, respectively), we have multimer predictions. Both 6OAN and 7MEZ in \autoref{fig:6oan} and \autoref{fig:7mez} each predict conformational ensemble of heterodimer structures with high accuracy. Especially the interface binding pose are well predicted and comparable with ground truth structures. 
Although there may be minor errors in predicted secondary structures not involved in interface binding, correct interface binding pose between proteins by non-bonded interactions is of greater importance. Minor misfolding may be addressed with methods such as MD, Monte Carlo and protein design 
tools~\cite{branden2021advances, maeots2022structural, amann2023frozen, schmidt2019time, martin2015comparing, leman2020macromolecular, sala2021insights, sala2019atomistic, matsunaga2020use, cerofolini2019integrative, allison2020computational, bussi2020using, webb2016comparative, kaufmann2010practically}. 

\begin{figure}[!tbhp]
\centering
\centerline{
\includegraphics[width=.25\textwidth, height=.25\textwidth]{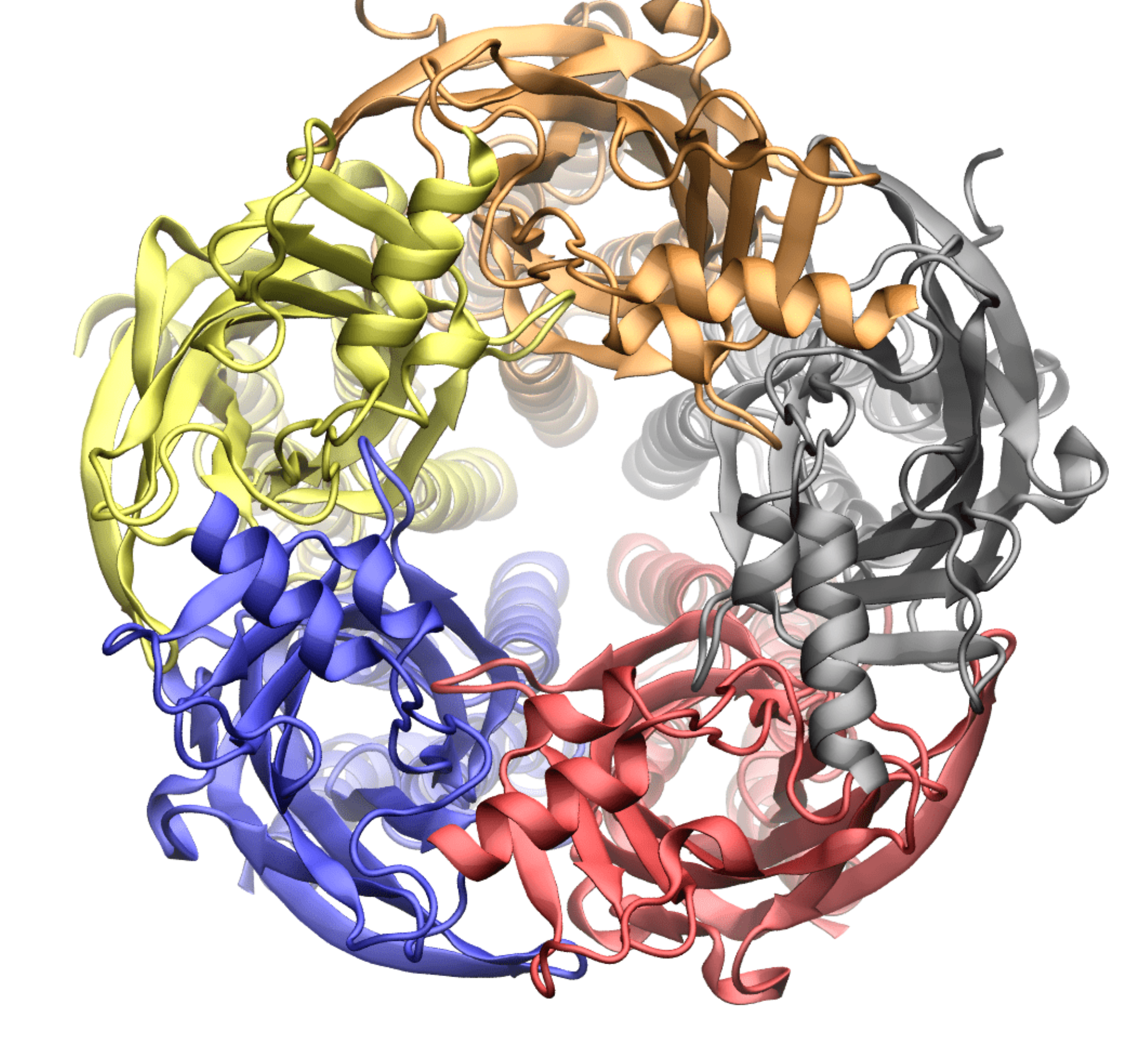}
\includegraphics[width=.25\textwidth, height=.25\textwidth]{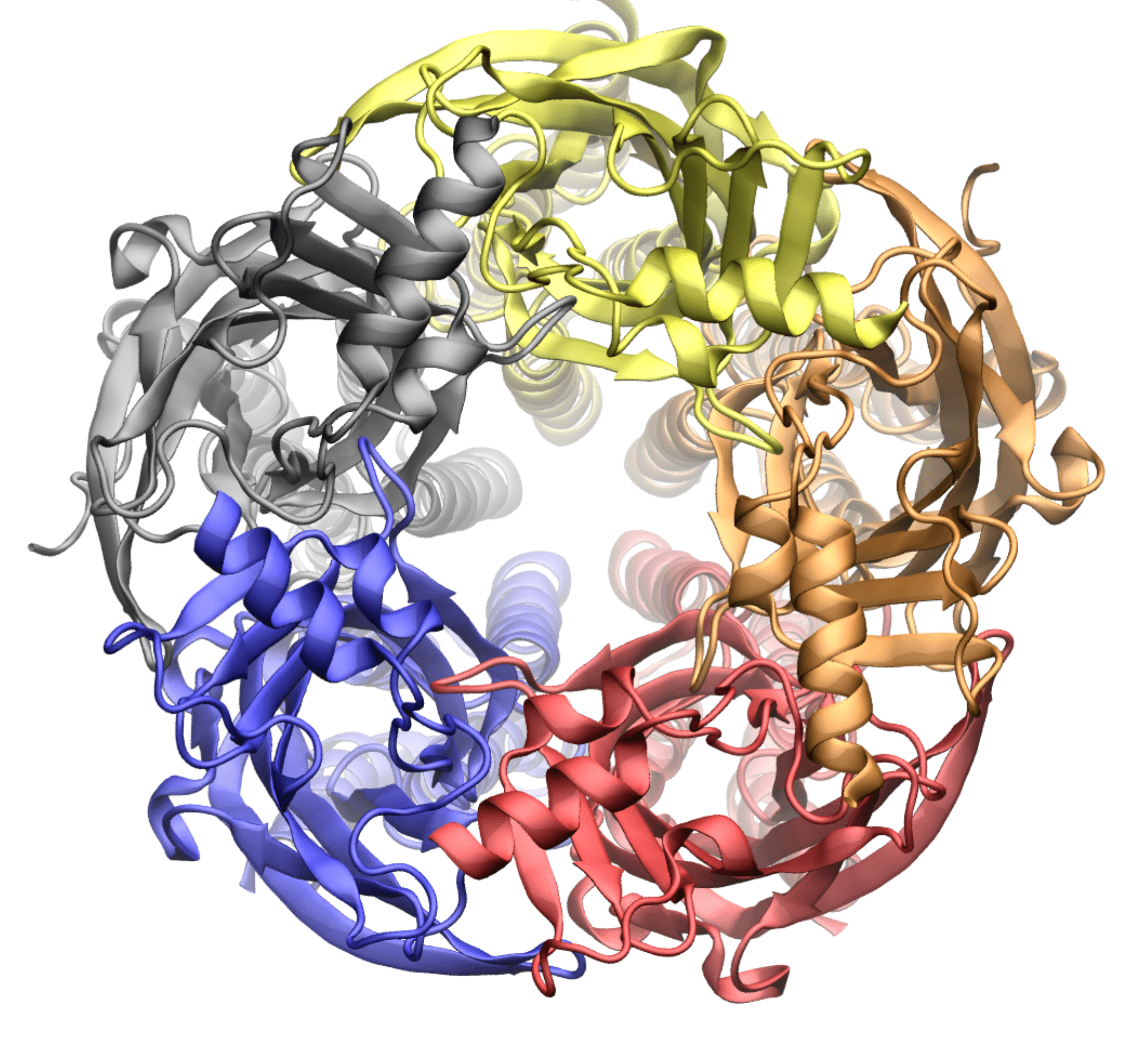}
}
\vspace{-3mm}
\caption{Protein structure used to test 
{\cal{\texttt{APACE}}}: a pentameric 
$\textrm{GABA}_{\textrm{A}}$ receptor (PDB 
accession: 6D6U). We show one predicted heteropentamer structure of neurotransmitter $\textrm{GABA}_{\textrm{A}}$ receptor. The left panel shows a comparable structure with ground truth predictions. Blue and gray chains form a homodimer while red and orange chains form the other homodimer. Yellow chain is a monomer differing in sequence from other two homodimers. However, the location of transmembrane helices (towards the paper direction) does not exactly reproduce the ground truth structure. This is understandable since \texttt{APACE} does not use membrane as an input to predict the transmembrane domain. However, the overall structure is comparable with the ground truth. On the other hand, in the right panel, we see an AI predicted protein whose structure is erroneous, and where blue and gray chains bind to each other. This structure may have high thermodynamic instability and steric hindrance when being crystallized.}
\label{fig:6d6u}
\end{figure}

In 6D6U in~\autoref{fig:6d6u}, we observe comparable structure (left panel) with ground truth and wrong structure (right panel) with wrong homodimer location predictions. Since 6D6U is a membrane petameric heteromer protein, it lends a challenging case of predicting not only correct structure of each monomer but also alternating chain patterns. The transmembrane helices are therefore mispredicted but overall structure is still comparable with ground truth.

In short, we have demonstrated \texttt{APACE}'s capabilities to 
predict protein structures, mirroring \texttt{AlphaFold2}'s robustness 
and accuracy, and providing remarkable speedups, reducing 
time-to-solution from days to minutes. \texttt{APACE} may be limited occasionally when it comes to predicting transmembrane proteins and/or multi-chain multimers, features it has inherited from \texttt{AlphaFold2}.

\section*{Methods}
Given that 
Delta and Polaris' container support is only available for 
Apptainer/Singularity~\cite{kurtzer2017singularity}, 
we modified the instructions provided in 
\texttt{AlphaFold2} GitHub repository, which 
are intended for Docker containers~\cite{merkel2014docker}. Below we 
describe the steps followed to deploy \texttt{AlphaFold2} on Delta and Polaris~\cite{towns2014xsede}:

\begin{enumerate}[nosep]
    \item We began by cloning the \texttt{AlphaFold2} directory from DeepMind and navigated to the respective directory. The code is available at \url{https://github.com/hyunp2/alphafold/tree/main}.
    \item Next, we downloaded the necessary genetic  databases for MSA capturing sequence representation, as well as for template similarity capturing structure representation. We also downloaded \texttt{AlphaFold2} model parameters required for \texttt{AlphaFold2}'s functioning. Every database and model parameter is up-to-date (as of August 2023)  and the multimer version is v3.
    \item Since Singularity is used on Delta and Polaris for container support, we built the Singularity image by initially building the Docker image locally. Afterward, we pushed this Docker image to DockerHub. Utilizing the \textit{singularity pull} command, we converted the Docker image to Singularity \textit{sif} format, making it compatible with Delta and Polaris environments. 
    
    Running the Singularity image built using the default Dockerfile resulted in an \texttt{HHsearch} (used for template search against PDB database) runtime error. To address this issue, modifications were made to the Dockerfile. Initially, the Dockerfile involved cloning and compiling the \texttt{HHsuite} package from source locally, which posed portability challenges across different machines. The compilation process with \textit{cmake} relied on the processor architecture of the user's machine, potentially leading to compatibility issues. For instance, if the user building the Docker image locally had a processor with an Instruction Set Architecture (ISA) that differed from Delta's supported architecture, \texttt{HHsearch} encountered a runtime error with ``Illegal Instruction''. 
    
    To address this issue and ensure cross-machine compatibility, we made modifications to the DockerFile. Instead of compiling \texttt{HHsuite} from source, we adopted a different approach by installing \texttt{HHsuite} using a statically compiled version that supports the AVX2 ISA. This modification eliminated the dependency on local processor architecture during the build process, mitigating the potential runtime errors and enhancing the portability of the Singularity image.
    
    \item To complete the setup, we created an output directory (default is \textit{/tmp/alphafold}) and ensured that it had the necessary permissions to allow writing.
    \item Due to Delta and Polaris' absence of Docker support, the standard \textit{run\_docker.py} script was not viable. Instead, we devised a custom shell script to replicate the essential functionality of \textit{run\_docker.py}. Employing a \textit{singularity run} command, we effectively bound the necessary mounts and passed the required flags for execution, mirroring the procedure of \textit{run\_alphafold.py} with Docker.
    \item Upon completion of the deployment process, the output directory contained the predicted structures of the target protein, accurately obtained through \texttt{AlphaFold2}'s advanced prediction capabilities.
\end{enumerate}

In conclusion, deploying \texttt{AlphaFold2} on Delta and Polaris required a series of modifications to account for Singularity containerization. Through this approach, we successfully integrated \texttt{AlphaFold2}'s powerful protein folding prediction capabilities into these supercomputers' environments.

\subsection{Limitations of \texttt{AlphaFold2} model}
\label{limits}

The original \texttt{AlphaFold2} model, while highly accurate in predicting protein structures, does have some limitations in terms of computational efficiency. Some of the key limitations include:

\begin{enumerate}[nosep]
    \item \textbf{Long Inference Time:} The time taken for the model to make predictions can be considerable, especially for larger and more complex protein structures. This can hinder its use in time-sensitive applications. Such long time inference including both CPU and GPU computations have been reported and analyzed 
    elsewhere in the literature~\cite{zhong2022parafold}.
    \item \textbf{Computationally Intensive/Limited Real-Time Predictions:} The original \texttt{AlphaFold2} model is computationally demanding, requiring significant computational resources and time for accurate predictions. For example, MSA and template search have to be performed in CPU  while GPU is utilized for each structure prediction, both in a sequential manner. This restricts its applicability for real-time predictions or on hardware with limited computational power. The computational demands of the model may hinder real-time prediction of protein structures, making it less suitable for time-sensitive applications.
    \item \textbf{Resource-Intensive:} \texttt{AlphaFold2}'s inference requires substantial computational resources, including powerful GPUs or tensor processing units (TPUs). This may limit its accessibility to researchers or institutions without access to high-end hardware. Also, the storage of database amounts to 2.6 TB, which may far exceed normal workstation storage capacity.
    \item \textbf{Memory Requirements:} The model's memory footprint can be substantial (with larger protein requiring higher GPU memory), making it challenging to process multiple protein structures concurrently, particularly on machines with limited RAM.
    \item \textbf{Single GPU Utilization:} The original \texttt{AlphaFold2} model is designed to use a single GPU during inference, limiting its capability to work with multiple GPUs. As a result, it predicts and relaxes one protein structure (saved as PDB file format) at a time sequentially.
    \item Other potential limitations include but are not limited to little protein conformation diversity. Predicting correct, yet diverse protein conformations, is a significant task for drug discovery, partially addressed in~\cite{faezov2023alphafold2, meller2023accelerating}.
\end{enumerate}

\subsection{Key optimizations in \texttt{APACE}}
\label{opt}

To transcend the limitations of \texttt{AlphaFold2}, we implemented optimizations in both CPU and GPU computation, striving for enhanced efficiency and performance.

\subsubsection{CPU optimization}
\texttt{AlphaFold2} utilizes \texttt{Jackhmmer} 
to conduct MSA searches on Uniref90, clustered MGnify and small BFD database. On the other hand, it employs \texttt{HHBlits} for MSA search on large BFD and Uniclust30 databases. In addition, \texttt{AlphaFold2} utilizes \texttt{HHSearch} for template search against PDB70 in the monomer case and \texttt{Hmmsearch} for template search against PDB Seqres in the multimer case. For the multimer only, \texttt{Jackhmmer} based Uniprot database parsing step exists as well.

To process a single query, \texttt{AlphaFold2} limits itself to 8 CPU cores for \texttt{Jackhmmer}, 4 CPU cores for \texttt{HHblits}, and 8 CPU cores for \texttt{Hmmsearch}. Given the vast database sizes (around 2.6 TB) and the considerable amount of I/O access involved, the MSA search for a single prediction can take several hours, significantly impacting the overall runtime (see~\autoref{limits} for bottlenecks for CPU computation).

To expedite the CPU stage, we implemented an approach inspired by ParaFold~\cite{zhong2022parafold}. By orchestrating the three independent sequential MSA searches in parallel, \texttt{APACE} significantly enhances the speed of MSA construction. In contrast to \texttt{AlphaFold2}, where UniRef90, MGnify, and BFD datasets were parsed sequentially using \texttt{Jackhmmer} and/or \texttt{HHblits} (for large BFD), \texttt{APACE} employed the Ray library~\cite{moritz2018ray} to simultaneously initiate three processes, enabling MSA searches to run concurrently.

Additionally, we allocated 16 CPU cores to each MSA search tool and template search tool: \texttt{Jackhmmer}, \texttt{HHblits}, and \texttt{HHSearch}/\texttt{Hmmsearch} (monomer/multimer). By running all three MSA tools in parallel, utilizing a total of 48 CPU cores, we achieved a substantial 1.8X speedup in performance.

After increasing the number of CPU cores, we observed remarkable speed enhancements in the MSA computation. However, beyond 20 CPU cores, the speed-up in MSA calculation plateaued. This observation unveiled the bottleneck as being related to input retrieval rather than CPU processing, rendering it input-bound. This inherent nature of being input-bound presents hurdles for straightforward parallelization (i.e., CPU multiprocessing or MPI) methods. Upon the completion of parallel computation for MSA, the template search and multimer Jackhmmer-Uniprot, sequentially ensue.

To further enhance speed in CPU intensive MSA computation, we made two key optimizations. Firstly, we migrated the entire dataset to solid state drives (SSD), minimizing data retrieval time. Additionally, we leveraged the Infinite Memory Engine (IME) to stage the dataset files into the SSD cache within the \textit{/ime} file system. This pre-staging allowed jobs to swiftly access and utilize the required data. IME is a DataDirect Networks (DDN) solution designed to facilitate fast data tiering between compute nodes and a file system within a HPC environment.

The MSA and structural template search results acquired on CPUs are stored in \textit{features.pkl} and passed to the neural network for prediction on GPUs. Additionally, we have incorporated a code check in our pipeline to circumvent CPU-burdening MSA computation. That is, if \textit{features.pkl} already exists (as a result of storing \textit{features.pkl} by successfully executing CPU computation at least once for a given protein sequence), the pipeline skips the MSA and structure template search and computation steps and proceeds directly to predict the protein structure, in a similar manner as~\cite{zhong2022parafold}. This optimization ensures efficient processing and avoids redundant computations.

\subsubsection{GPU optimization}

\texttt{AlphaFold2}/\texttt{APACE} employs an ensemble of five neural network models to predict the 3D structure of proteins. This ensemble approach entails using multiple pretrained models with slight variations hyperparameters for protein structure prediction. The three models out of five make predictions based on MSA (i.e., models 3-5) while the other two models (i.e., models 1-2) also rely on templates. For details of how the five models differ, we refer the refer to~\cite{jumper2021highly, sala2023modeling} 

The \textit{``-{}-num\_multimer\_predictions\_per\_model"} flag governs the number of independent predictions made by each individual neural network model within the model ensemble. When running \texttt{AlphaFold2}/\texttt{APACE}, users can specify the value for this flag, thereby controlling the number of predictions generated by each model. The collective predictions from each model in the model ensemble offer diverse final predicted 3D structures (e.g., plasmepsin II, an aspartic protease causing malaria \cite{meller2023accelerating}), which are crucial to understand free energy landscape of protein conformations and to identify important drug discovery target, cryptic binding pockets. In contrast to \texttt{AlphaFold2}, we included in \texttt{APACE} a capability to predict multiple monomer structures per model. 

The original \texttt{AlphaFold2} model is designed to use a single GPU during inference, which does not take full advantage of deep learning's parallel processing capabilities. During GPU utilization, \texttt{AlphaFold2} performs sequential structure prediction, which is one of the reasons why \texttt{AlphaFold2} takes a long time till completion~\cite{zhong2022parafold} (see \autoref{limits}). To expedite the GPU phase, we used the Ray library for GPU parallelization as well for \texttt{APACE}. Therefore, each ensemble model and its corresponding predictions are allocated to distinct GPUs for structure prediction. As a result, \texttt{APACE} can harness multiple GPUs to efficiently run models in parallel, markedly expediting the overall prediction process.

Following prediction by each model, the corresponding structure undergoes a relaxation process in a sequential manner in \texttt{AlphaFold2}. To enhance the efficiency of this step, we once more harnessed the power of the Ray library in \texttt{APACE}. Through this optimization, each structure predicted by ensemble of models is assigned to an individual dedicated GPU, facilitating parallel relaxation processing. This enhancement by \texttt{APACE} has substantially reduced processing time, contributing to the overall acceleration of the relaxation process.

\section*{Conclusions}
\label{conclusions}

We have introduced \texttt{APACE}, a framework that 
retains the robustness and accuracy of \texttt{AlphaFold2}, 
and which leverages supercomputing to reduce 
time-to-insight from days to minutes. We have accomplished 
this by a) making an efficient use of the 
Delta and Polaris supercomputers systems' data storage and data 
staging; b) optimizing CPU and GPU computing; and c) developing 
scientific software to enable the prediction of conformational ensemble of protein structures. 
These tools are released with this 
manuscript to provide researchers with a 
computational framework 
that may be readily linked with robotic laboratories to 
automate and accelerate scientific discovery.

\showmatmethods{} 

\section*{Data and Software Availability}
The data and scientific software needed to reproduce 
this work is available at: \url{https://github.com/hyunp2/alphafold/tree/main}. 

\acknow{This work was supported by Laboratory 
Directed Research and Development (LDRD) funding 
from Argonne National Laboratory, provided by the 
Director, Office 
of Science, of the U.S. Department of Energy under 
Contract No. DE-AC02-06CH11357. E.A.H. was 
partially supported by National Science Foundation 
award OAC-2209892. This research 
used resources of the Argonne Leadership Computing 
Facility, which is a DOE Office of Science User 
Facility supported under Contract DE-AC02-06CH11357. 
This research used the Delta advanced computing and 
data resource which is supported by the National 
Science Foundation (award 
OAC 2005572) and the State of Illinois. Delta is a 
joint effort of the University of Illinois at
Urbana-Champaign and its National Center for 
Supercomputing Applications.}

\showacknow{} 



\bibliography{pnas-sample}

\begin{thebibliography}{10}

\bibitem{DeepLearning}
LeCun Y, Bengio Y, Hinton G (2015) Deep learning.
\newblock {\em Nature} 521(7553):436--444.

\bibitem{Nat_Rev_2019_Huerta}
{Huerta} EA, et~al. (2019) {Enabling real-time multi-messenger astrophysics
  discoveries with deep learning}.
\newblock {\em Nature Reviews Physics} 1(10):600--608.

\bibitem{2022NatRP...4..761K}
{Krenn} M, et~al. (2022) {On scientific understanding with artificial
  intelligence}.
\newblock {\em Nature Reviews Physics} 4(12):761--769.

\bibitem{sci_miw}
{Special Issue} (2023) A machine-intelligent world.
\newblock {\em Science} 381(6654):136--137.

\bibitem{BIANCHINI2022104604}
Bianchini S, Müller M, Pelletier P (2022) Artificial intelligence in science:
  An emerging general method of invention.
\newblock {\em Research Policy} 51(10):104604.

\bibitem{Crawford+2021}
Crawford K (2021) {\em Atlas of AI: Power, Politics, and the Planetary Costs of
  Artificial Intelligence}.
\newblock (Yale University Press, New Haven).

\bibitem{dl2010review}
Dean J (2022) {A Golden Decade of Deep Learning: Computing Systems \&
  Applications}.
\newblock {\em Daedalus} 151(2):58--74.

\bibitem{openai2023gpt4}
OpenAI (2023) Gpt-4 technical report.

\bibitem{jumper2021highly}
Jumper J, et~al. (2021) Highly accurate protein structure prediction with
  alphafold.
\newblock {\em Nature} 596(7873):583--589.

\bibitem{bryant2022improved}
Bryant P, Pozzati G, Elofsson A (2022) Improved prediction of protein-protein
  interactions using alphafold2.
\newblock {\em Nature communications} 13(1):1265.

\bibitem{moult2020critical}
Moult J, Fidelis K, Kryshtafovych A, Schwede T, Topf M (2020) Critical
  assessment of techniques for protein structure prediction, fourteenth round.
\newblock {\em CASP 14 Abstract Book}.

\bibitem{evans2021protein}
Evans R, et~al. (2021) Protein complex prediction with alphafold-multimer.
\newblock {\em biorxiv} pp. 2021--10.

\bibitem{zhong2022parafold}
Zhong B, et~al. (2022) Parafold: paralleling alphafold for large-scale
  predictions in {\em International Conference on High Performance Computing in
  Asia-Pacific Region Workshops}.
\newblock pp. 1--9.

\bibitem{mirdita2022colabfold}
Mirdita M, et~al. (2022) Colabfold: making protein folding accessible to all.
\newblock {\em Nature methods} 19(6):679--682.

\bibitem{yin2022benchmarking}
Yin R, Feng BY, Varshney A, Pierce BG (2022) Benchmarking alphafold for protein
  complex modeling reveals accuracy determinants.
\newblock {\em Protein Science} 31(8):e4379.

\bibitem{meller2023accelerating}
Meller A, Bhakat S, Solieva S, Bowman GR (2023) Accelerating cryptic pocket
  discovery using alphafold.
\newblock {\em Journal of Chemical Theory and Computation}.

\bibitem{faezov2023alphafold2}
Faezov B, Dunbrack~Jr RL (2023) Alphafold2 models of the active form of all 437
  catalytically-competent typical human kinase domains.
\newblock {\em bioRxiv} pp. 2023--07.

\bibitem{sala2023modeling}
Sala D, Engelberger F, Mchaourab H, Meiler J (2023) Modeling conformational
  states of proteins with alphafold.
\newblock {\em Current Opinion in Structural Biology} 81:102645.

\bibitem{towns2014xsede}
Towns J, et~al. (2014) Xsede: accelerating scientific discovery.
\newblock {\em Computing in science \& engineering} 16(5):62--74.

\bibitem{moritz2018ray}
Moritz P, et~al. (2018) Ray: A distributed framework for emerging $\{$AI$\}$
  applications in {\em 13th USENIX symposium on operating systems design and
  implementation (OSDI 18)}.
\newblock pp. 561--577.

\bibitem{humphrey1996vmd}
Humphrey W, Dalke A, Schulten K (1996) Vmd: visual molecular dynamics.
\newblock {\em Journal of molecular graphics} 14(1):33--38.

\bibitem{coleman2019serotonin}
Coleman JA, et~al. (2019) Serotonin transporter--ibogaine complexes illuminate
  mechanisms of inhibition and transport.
\newblock {\em Nature} 569(7754):141--145.

\bibitem{chan2022structural}
Chan MC, Procko E, Shukla D (2022) Structural rearrangement of the serotonin
  transporter intracellular gate induced by thr276 phosphorylation.
\newblock {\em ACS Chemical Neuroscience} 13(7):933--945.

\bibitem{chan2022substrate}
Chan MC, Selvam B, Young HJ, Procko E, Shukla D (2022) The substrate import
  mechanism of the human serotonin transporter.
\newblock {\em Biophysical journal} 121(5):715--730.

\bibitem{branden2021advances}
Br{\"a}nd{\'e}n G, Neutze R (2021) Advances and challenges in time-resolved
  macromolecular crystallography.
\newblock {\em Science} 373(6558):eaba0954.

\bibitem{maeots2022structural}
M{\"a}eots ME, Enchev RI (2022) Structural dynamics: Review of time-resolved
  cryo-em.
\newblock {\em Acta Crystallographica Section D: Structural Biology} 78(8).

\bibitem{amann2023frozen}
Amann SJ, Keihsler D, Bodrug T, Brown NG, Haselbach D (2023) Frozen in time:
  analyzing molecular dynamics with time-resolved cryo-em.
\newblock {\em Structure}.

\bibitem{schmidt2019time}
Schmidt M (2019) Time-resolved macromolecular crystallography at pulsed x-ray
  sources.
\newblock {\em International Journal of Molecular Sciences} 20(6):1401.

\bibitem{martin2015comparing}
Mart{\'\i}n-Garc{\'\i}a F, Papaleo E, Gomez-Puertas P, Boomsma W,
  Lindorff-Larsen K (2015) Comparing molecular dynamics force fields in the
  essential subspace.
\newblock {\em PLoS One} 10(3):e0121114.

\bibitem{leman2020macromolecular}
Leman JK, et~al. (2020) Macromolecular modeling and design in rosetta: recent
  methods and frameworks.
\newblock {\em Nature methods} 17(7):665--680.

\bibitem{sala2021insights}
Sala D, Giachetti A, Rosato A (2021) Insights into the dynamics of the human
  zinc transporter znt8 by md simulations.
\newblock {\em Journal of Chemical Information and Modeling} 61(2):901--912.

\bibitem{sala2019atomistic}
Sala D, Giachetti A, Rosato A (2019) An atomistic view of the yiip structural
  changes upon zinc (ii) binding.
\newblock {\em Biochimica et Biophysica Acta (BBA)-General Subjects}
  1863(10):1560--1567.

\bibitem{matsunaga2020use}
Matsunaga Y, Sugita Y (2020) Use of single-molecule time-series data for
  refining conformational dynamics in molecular simulations.
\newblock {\em Current opinion in structural biology} 61:153--159.

\bibitem{cerofolini2019integrative}
Cerofolini L, et~al. (2019) Integrative approaches in structural biology: a
  more complete picture from the combination of individual techniques.
\newblock {\em Biomolecules} 9(8):370.

\bibitem{allison2020computational}
Allison JR (2020) Computational methods for exploring protein conformations.
\newblock {\em Biochemical Society Transactions} 48(4):1707--1724.

\bibitem{bussi2020using}
Bussi G, Laio A (2020) Using metadynamics to explore complex free-energy
  landscapes.
\newblock {\em Nature Reviews Physics} 2(4):200--212.

\bibitem{webb2016comparative}
Webb B, Sali A (2016) Comparative protein structure modeling using modeller.
\newblock {\em Current protocols in bioinformatics} 54(1):5--6.

\bibitem{kaufmann2010practically}
Kaufmann KW, Lemmon GH, DeLuca SL, Sheehan JH, Meiler J (2010) Practically
  useful: what the rosetta protein modeling suite can do for you.
\newblock {\em Biochemistry} 49(14):2987--2998.

\bibitem{kurtzer2017singularity}
Kurtzer GM, Sochat V, Bauer MW (2017) Singularity: Scientific containers for
  mobility of compute.
\newblock {\em PloS one} 12(5):e0177459.

\bibitem{merkel2014docker}
Merkel D (2014) Docker: lightweight linux containers for consistent development
  and deployment.
\newblock {\em Linux journal} 2014(239):2.

\end{thebibliography}

\end{document}